\documentstyle[12pt]{article}
\begin{document}
\def\be{\begin{equation}}
\def\ee{\end{equation}}
\def\ba{\begin{eqnarray}}
\def\ea{\end{eqnarray}}

\begin{center}
{\large\bf Unitarity Effects at Low x}\\

\vspace*{0.2cm}
{\bf Vladimir A. Petrov} \\
\small
{Division of Theoretical Physics, Institute for High Energy Physics,
142284 Protvino, 
Russian Federation} \\
\end{center}

\vspace*{0.5cm}
\small
{We discuss effects of unitarity on the low-x behaviour of the structure
functions of deeply  inelastic scattering. Particular attention is paid to
the distinction between Regge  and Renormalization Group $J$-plane
singularities and their relative weight in defining the low-x asymptotics.}

\vspace*{0.8cm}
As is well known, unitarity (+ some other basic  principles) leads to a quite
severe restriction of  possible growth of the hadronic cross-sections at
asymptotically high energies, $\sqrt{s}$, [1]:
\be
\sigma^{hh}_{tot}(s)\leq
\frac{\pi}{m^2_\pi}(\log s)^2,\;\; s\to\infty.
\ee
At present energies the Froissart-Martin upper bound (1) is not very actual, so
that even powerlike growth (formally more rapid then (1))
\be
\sigma^{hh}_{tot}(s)\sim s^\Delta
\ee
(with $\Delta$ typically of order 0.1) is fairly admissible as a transient,
pre-asymptotic behaviour [2]. The mechanisms transforming (2)
into (1) at larger energies ("unitarization" ) are given e.g. by (various versions
of) the Regge-Eikonal approach [3] or $U$-Matrix approach [4], both taking use
of the leading Regge trajectory $\alpha (t)$ with $\alpha (0)=1+\Delta,\;\;
\Delta>0$. 

At very high energies (2) is converted into 
\be
\sigma^{hh}_{tot}\simeq 8\pi\alpha'(0)\Delta\log^2s+...
\ee
with a (very weak) condition $8\alpha'(0)\Delta\leq m_\pi^{-2}$. 
The importance of the "unitarity corrections" depends strongly on the value of
$\Delta$, which still is not calculated theoretically.

Similar questions can arise in deeply inelastic electron-proton scattering.  In
particular it was reported recently [5] that a powerlike growth
\be
\sigma^{\gamma^*p}_{tot}(s,Q^2)\sim
\left(\frac{s}{Q^2}\right)^\lambda\approx
\left(\frac{1}{x}\right)^\lambda
\ee
at $s>>Q^2\geq(\mbox{hadronic mass scale})^2$, fits well the data with 
$$
0.2(Q^2\simeq 1\mbox{GeV}^2)\leq\lambda\leq 0.5(Q^2\simeq 10^3\mbox{GeV}^2).
$$
As the exponent $\lambda$ from Eq.(4) is significantly larger than $\Delta$
from Eq.(2) one could think that "unitarity corrections" should be more
important for deep-inelastic scattering and the bound (1) is more restrictive. 

However as was indicated already many years ago in Ref.[6], the
Froissart-Martin bound cannot be proven for off-mass-shell amplitudes, what
is the case for deeply inelastic scattering. Moreover, it is impossible to do
that rigorously for any current-hadron amplitude [7].
Some recent attempts [8,9] to preserve the bound (1) by the price of new assumptions
do not possess a general power inherent to the Froissart-Martin theorem.

Nonetheless, unitarity influences high energy behaviour of the off-shell
amplitude as well [7]. 
To exhibit explicitly the unitary effects 
we take use of the Regge-Eikonal approach. 
As was shown in Ref.[7] the amplitude describing deeply inelastic scattering 
in the impact parameter space may be presented in the following form (below we
consider for simplicity all particles as scalars):
$$
T^{**}(s,b,q^2)=\delta^{**}(s,b,q^2)-
[\delta^*]^2/\delta+[\delta^*/\delta]^2T(s,b),
$$
where asterisks denote the number of the off-shell particles,
$\delta$ and $\delta^*,\delta^{**}$ stand for the eikonal and its off-shell
extentions. Remind that 
$$
T=\frac{i}{2}[1-e^{2i\delta}]
$$
and
$$
Im\delta\geq0,\;\; \mbox{at}\;\; s\geq s_{{\small\mbox{inelastic}}}.
$$
One can obtain from unitarity that
\be
Im(\delta^{**}-[\delta^*]^2/\delta)\geq
(Im(\delta^{*}/\delta))^2|T|^2/ImT.
\ee
If we neglect real parts of eikonal functions we get from (5) 
$(\Omega\equiv Im\delta)$
\be
\Omega^{**}\Omega\geq (\Omega^*)^2.
\ee
In the framework of the Regge-Eikonal approach
\be
\Omega(s,b)\approx \frac{c\cdot s^\Delta}{\pi \rho^2(s)}e^{-b^2/\rho^2(s)},
\ee
where $c=$constant, $\rho^2(s)\simeq 4\alpha'(0)\log s+...$. To proceed further we
identify $\Omega^{**}(\Omega^*)$ with the twist-2 contribution in the operator
product expansion of the amplitudes
$$
T^{**}(P,Q,\Delta)=\int dx^4e^{i\frac{q+q'}{2}x}<p'|
\frac{\delta J(-\frac{x}{2})}{\delta a(\frac{x}{2})}|p>
$$
and
$$
T^{**}(P,Q,\Delta)=\int dx^4e^{i\frac{q+q'}{2}x}<p'|
\frac{\delta J(-\frac{x}{2})}{\delta v(\frac{x}{2})}|p>,
$$
where $Q=(q'+q)/2,\;\; P=(p'+p)/2,\;\;
\Delta=q-q', \;\;J$ is the current, $a$ is the "electromagnetic" field and $v$ is
the "vector meson" field. In more detail this relation has the form
$$
\Omega^{**}(s,b,Q^2)=\int dt J_0(b\sqrt{-t})
\int\frac{dJ}{2i}\left(\frac{1}{x}\right)^J\cdot
$$
\be
\cdot C^{*(*)}_{J,i}
(\alpha_s(Q^2))f_{J,i}(Q^2,t),
\ee
where $C^{*(*)}_{J,i}$ is the coefficient function, and $f_J(Q^2,t)$ is the leading
(at small $t$) formfactor of the twist-two composite operator  
$O^i_{\mu_{1}...\mu_J}$ renormalized at $Q^2(t\equiv \Delta^2)$. Formfactor
$f_{J,i}(Q^2,t)$ may be interpreted as a following  double integral transform
of the parton density, $f^i(x,\vec b;Q^2)$ in the momentum fraction $x$ and
"transverse distance $\vec b$  from the center of the hadron" 
$$
f_{J,i}(Q^2,t)=
\int^1_0dx\,x^{J-1}\int d^2bJ_0
(b\sqrt{-t})\cdot
$$
\be
\cdot f_i(x,\vec b;Q^2).
\ee
Identification of $\delta's$ with the twist-2 contribution of the OPE can be
considered as a modern variant of the Chou-Yang interpretation [10] of the
eikonal representation.
Formfactors $f_{J,i}(Q^2,t)$ obey the Renormalization Group equation
$$
Q^2\frac{\partial}{\partial Q^2}f_{J,i}(Q^2,t)=
[\gamma_J(\alpha_s(Q^2))]_{ij}f_{J,j}(Q^2,t),
$$
where $\gamma_J$ is the anomalous dimension matrix.

It follows from the very general considerations that $f_J$ should contain
Regge poles with quantum numbers of the $t$-channel
$$
f_J\sim (J-\alpha (t))^{-1}.
$$
When all particles are on-shell these poles 
define the behaviour (7) of the
eikonal, and, eventually, asymptotic behaviour (3) of the total cross-secton.

Situation is more complicated in the off-shell case. It is well known that
anomalous dimensions in singlet channels develop singular behaviour at $J=1$
[11]; so that $\gamma_J(\alpha_s)$ is a formal power series in $\alpha_s
(Q^2)/(J-1)$, which diverges at small $(J-1)$ enough. 
There exists  a believe
[12] that resummation of this formal series results in a  "shifted" 
singularity at $J=1+\lambda,\;\;
\lambda>0$. In general $\lambda$ should  depend on $Q^2$.

We are now with two types of singularities: Regge singularities, $\alpha(t)$,
which are, naturally, $Q^2$-independent; and RG (Renormalization Group)
singularities, which depend on $Q^2$, but do not depend on $t$, due to the
multiplicative character of renormalization of composite operators. Both lie supposedly to the right of $ReJ=1$. Physical interpretation of the RG
singularities generated by anomalous dimensions   is not yet clear (contrary to the Regge
singularities). It seems that for the on-mass-shell amplitudes they should be
absent if we hold to the ideology of "maximal analyticity of the second degree" [13].
One can compare this situation with Ref.[14] where the question on "two
pomerons" was addressed on purely phenomenological basis. 
Another attempt of a phenomenological account of unitarity in DIS was
undertaken in Ref.[15].

It is naturally to address the question: what singularity dominates, i.e., is
the rightmost one in the $J$-plane?  From what was said about observed
behaviour of the structure functions at small $x$ a temptation  
arises 
to conclude
that the structure functions of DIS are dominated by RG singularities at $J=1+\lambda(Q^2)$.

It is not evident, however, that this scenario is compatible with general
condition (5).

Neglect for a moment the real parts of eikonals $\delta, \delta^*,\delta^{**}$
and
assume that $\lambda (Q^2)>\Delta =\alpha(0)-1$. Integration of Eq.(6) in
$b$ leads to the following inequality:
\be
a(Q^2)\left(\frac{1}{x}\right)^{\lambda (Q^2)}\leq
\left(\frac{1}{x}\right)^{\Delta},\;\; x<<1,
\ee
where $a(Q^2)$ does not depend on $x$. It's clear, that $\lambda (Q^2)$ cannot
be larger than $\Delta$. Otherwise (10) will be violated for $x$ small enough. It
is, however, possible that in some limited intervals of $x$ and $Q^2$  one could
observe a faster rise of $\sigma^{\gamma^*p}_{tot}$ with energy at larger
$Q^2$ as some transient phenomenon. But at $1/x\to \infty$  the
"off-shell" cross-sections will have $(1/x)^\Delta$ behaviour with 
a universal 
"soft" exponent
$\Delta=\alpha (0)-1$. 

\section*{Acknowledgments}
I am grateful to Stephan Narison for inviting me to QCD-96.

\end{document}